\documentstyle[aps,preprint]{revtex}

\begin{document}

\newcommand{\bqq}{\begin{equation}}
\newcommand{\eqq}{\end{equation}}

\title{Orthorhombic Phase of Crystalline Polyethylene: A Monte Carlo Study}

\author{R. Marto\v{n}\'ak$^{(1,2,}$\cite{byline}$^)$, W. Paul$^{(1)}$,
K. Binder$^{(1)}$}

\address{
$^{(1)}$ Institut f\"ur Physik, KoMa 331, Johannes Gutenberg--Universit\"at \\
Staudingerweg~7, 55099~Mainz, Germany \\
$^{(2)}$ Max--Planck--Institut f\"ur Polymerforschung \\
Ackermannweg~10, 55021~Mainz, Germany \\}

\date{29 November 1996}
\maketitle

\begin{abstract}
\noindent
In this paper we present a classical Monte Carlo simulation of the orthorhombic
phase of crystalline polyethylene, using an explicit atom force field with 
unconstrained bond lengths and angles and periodic boundary conditions. 
We used a recently developed
algorithm which apart from standard Metropolis local moves employs also
global moves consisting of displacements of the center of mass of the whole 
chains in all three spatial directions as well as rotations of the chains 
around an axis parallel to the crystallographic $c$-direction. Our simulations 
are performed in the $NpT$ ensemble, at zero pressure, and extend over the whole
range of temperatures in which the orthorhombic phase is experimentally known 
to be stable (10 -- 450 K). In order to investigate the finite-size effects in 
this extremely anisotropic crystal, we used different system sizes and different
chain lengths, ranging from C$_{12}$ to C$_{96}$ chains, the total number of 
atoms in the super-cell being between 432 and 3456. We show here the results for 
structural parameters, such as the orthorhombic cell parameters $a,b,c$, and 
the setting angle of the chains, as well as internal parameters of the
chains, such as the bond lengths and angles. Among thermodynamic quantities,
we present results for thermal expansion coefficients, elastic constants and 
specific heat. We discuss the temperature dependence of the measured quantities 
as well as the related finite-size effects. In case of lattice parameters and
thermal expansion coefficients, we compare our results to those obtained from other 
theoretical approaches as well as to some available experimental data. We also
suggest some possible ways of extending this study.

\end{abstract}

\draft

\section{Introduction}

Macromolecular crystals represent at the same time a particular case of a solid 
state system as well as that of a polymer system. The simplest and paradigmatic 
case in this field is crystalline polyethylene. As is well known, it is very
difficult to obtain reliable experimental data for these systems, mainly due
to problems associated with preparing sufficiently large monocrystalline 
samples. This increases the value of theoretical insight for understanding and 
eventual prediction of their properties. However, compared to the case of 
liquid polymer phases, a quantitative description of structure and properties 
of a solid phase is a more subtle problem. While 
in the former case it is often possible to replace whole monomer groups 
by united atoms and constrain the bond lengths and sometimes even the bond 
angles, in case of macromolecular crystals, instead, it is preferred to employ
a force field for all atoms, without enforcing any constraints on the degrees 
of freedom \cite{wunderlich}.

Apart from the study of the ground state crystal structure and properties, 
which can be readily done within the framework of molecular mechanics, once 
a force field is available, the main interest lies in the finite temperature 
properties. At low temperatures, where the creation of conformational defects 
is unlikely, and the displacements of the atoms are relatively small,
the structure of the PE crystal is dominantly governed by packing energetics of 
all-trans conformation chains. In this regime, phonon modes are well defined
excitations of the system, and a possible theoretical approach is the use of 
quasi-harmonic or self-consistent quasi-harmonic approximations, which have 
the advantage of allowing for quantum effects to be taken into account 
easily \cite{rutledge,hagele,stobbe}.

Such methods are, however, intrinsically incapable of treating correctly the
large amplitude, anharmonic motion of whole chain segments, which arises 
as the temperature approaches the melting point (at normal pressure about 
414 K for a PE crystal). Actually, for crystalline PE they start to 
fail just at the room temperature of 300 K \cite{stobbe}. If the melting is 
prevented by increasing the external pressure, the orthorhombic crystal 
structure may persist to higher temperatures, and eventually undergo a phase 
transition into the hexagonal "condis" phase, characterized by a large 
population of gauche defects \cite{condis}. In order to study the complex 
behavior in this high temperature regime, computer simulation techniques, like
molecular dynamics or Monte Carlo, appear as a particularly convenient tool. 
Using a suitable ensemble and an appropriate simulation technique one can 
evaluate various thermodynamic quantities, like specific heat, elastic constants
or thermal expansion coefficients, and structural quantities, such as bond 
lengths and angles, or defect concentration. It is also possible to directly 
study a variety of physically interesting phenomena associated with the 
structural phase transitions between different crystal modifications.

Computer simulations are, however, still limited to relatively small systems.
In case of macromolecular crystals, the situation is rather peculiar due to 
the extreme anisotropy of the system originating from its quasi-one-dimensional
character. When the chains are short, the system crosses over from a PE crystal
to an
$n$-paraffin crystal, which behaves in a substantially different way. Due to an
easy activation of chain rotation and diffusion, $n$-paraffin crystals undergo, 
as the temperature is increased, a characteristic series of phase transitions,
depending on the chain length ("rotator phases",\cite{ungar,rk}).
This fact makes it necessary to study and understand the related finite-size 
effects, if the results of the simulation are to be regarded as representative 
for the limit of very long chains, corresponding to PE, and interpreted in 
a consistent way. The choice
of boundary conditions also represents a non-trivial issue, as documented by 
the very extensive explicit atom MD simulation \cite{wunderlich}, in which 
a transition from an initial orthorhombic structure to a parallel zig-zag chain
arrangement was observed already at 111 K, contrary to experiment, probably
because of the use of free boundary conditions in all spatial directions and 
a related large surface effect. 
 
Finite temperature simulations of realistic, explicit atom models of crystals 
with long methylene chains, of which we are aware so far, have exclusively used
molecular dynamics as the sampling method. Ryckaert and Klein \cite{rk} used a 
version of the Parrinello-Rahman MD technique with variable cell shape to 
simulate an $n$-paraffin crystal with constrained bond lengths and periodic 
boundary conditions. Sumpter, Noid, Liang and Wunderlich have simulated 
large crystallites consisting of up to $10^4$ CH$_2$ groups, using free 
boundary conditions \cite{wunderlich}. 
Recently, Gusev, Zehnder and Suter \cite{gusev} have simulated PE 
crystal using periodic boundary conditions, at zero pressure, using also the 
Parrinello-Rahman technique.

On the other hand, not much MC work seems to exist in this field, and to our 
knowledge, the only work is that of Yamamoto \cite{yamamoto} which was aimed
specifically at the high temperature "rotator" phases of $n$-paraffins.
It concentrated exclusively on the degrees of freedom associated with the 
packing of whole chains, assumed to be rigid, neglecting completely the 
internal, intramolecular degrees of freedom.

The aim of this paper is twofold. On the one hand, we would like to explore the 
applicability of the MC method to a classical simulation of a realistic model of
a PE crystal, using an explicit atom force field without any constraints, and 
periodic boundary conditions in all spatial directions. In order to have 
a direct access to quantities 
like thermal expansion coefficients, or elastic constants, we choose to work
at constant pressure. If MC turns out to be a well applicable method
for this system on the classical level, it would be a promising sign for an 
eventual Path Integral Monte Carlo (PIMC) study allowing to take into account 
also the quantum effects, known to play a crucial role at low temperatures 
\cite{rutledge,hagele,stobbe}. It is known that for path integral techniques 
the use of MC is generally preferred to MD, because of the problem of 
non-ergodicity of the pseudo-classical system representing the quantum one in
a path integral scheme. As a related problem, we would also like to understand 
the finite-size effects and determine for each particular temperature
the minimal size of the system that has to be simulated in order to be 
reasonably representative of the classical PE crystal. Such information would 
also be very useful for an eventual PIMC study, where an additional
finite-size scaling has to be performed, because of the finite Trotter number.
On the other hand, we would like to study the physics of the orthorhombic phase 
of PE crystal in the classical limit over the whole temperature range of its
experimentally known existence at normal pressure. 
We stress that our goal here is not to tune the force field
in order to improve the agreement between the simulation and experiment. Our 
emphasis rather lies on providing results calculated for a given force field 
with an essentially exact classical technique. These might in future serve as 
a basis for estimation of the true importance of quantum effects, when these 
can be taken into account by a PIMC technique, as well as for an assessment of 
the range of validity of different classical and quantum approximation schemes. 
Some preliminary simulation results of this study have been presented briefly 
in recent conference proceedings \cite{rm1}.

The paper is organized as follows. In Sect.~II, we describe the force field 
as well as the constant pressure simulation method used.
The MC algorithm itself will be addressed only briefly, since it has already 
been discussed in detail in its constant volume version in Refs.\cite{rm,rm1}.
In Sect.~III, we present the results for the structural and 
thermodynamic properties of the orthorhombic PE crystal obtained from zero 
pressure simulation in the temperature range 10 -- 450 K, for different chain 
lengths and system sizes. We discuss the temperature dependence 
of the measured quantities as well as the related finite-size effects, and 
compare our results to those obtained from other theoretical approaches, as 
well as to some available experimental data.
In the final Sect.~IV we then draw conclusions and suggest some possible 
further directions.

\section{Constant pressure simulation method}

In this section we describe some details of the simulation method, as well as 
of the force field used. Before doing so, we recall here the structure of the 
orthorhombic PE crystal \cite{kavesh}. The unit cell contains two chains, each
consisting of 2 CH$_2$ groups, giving a total of 12 atoms per unit cell. The 
all-trans chains extend along the crystallographic $c$-direction ($z$-axis) 
and are packed in a "herringbone" arrangement, characterized by the setting angle 
$\psi$ (angle between the $xz$ plane and the plane containing the carbon backbone 
of a chain) alternating from one raw of chains to another between the values 
$\pm |\psi|$. The packing is completely
determined by specifying the 3 lattice parameters $a,b,c$ as well as the value of
$|\psi|$. To specify the internal structure of the chains, three 
additional parameters are needed, which may be taken to be the bond lengths 
$r_{CH}$ and $r_{CC}$ and the angle $\theta_{HCH}$. 

We have simulated a super-cell containing $i \times j \times k$ unit cells of
the crystal, $i,j,k$ being integers. Periodic boundary conditions were applied 
in all three spatial directions, in order to avoid surface effects. Our PE 
chains with backbones consisting of $2k$ carbon atoms are thus periodically 
continued beyond the simulation box and do not have any chain ends. 

As the study was aimed specifically at the orthorhombic phase of the PE crystal
and
we did not expect phase transitions into different crystal structures, we did 
not consider general fluctuations of the super-cell shape, otherwise common in 
the Parrinello - Rahman MD method. We have constrained the crystal structure 
angles $\alpha,\beta,\gamma$ to the right angle value, not allowing for shear
fluctuations. The volume moves employed thus consisted only of an anisotropic 
rescaling of the linear dimensions of the system by three scaling factors 
$s_1,s_2,s_3$, which relate the instantaneous size of the super-cell to that of
the reference one. The reference super-cell always corresponded to lattice 
parameters $a = 7.25 \AA, b = 5.00 \AA, c = 2.53 \AA$. The acceptance criterion
for the volume moves was based on the Boltzmann factor $(s_1 s_2 s_3)^N 
e^{-\beta H}$, where $H = U + p V_0 s_1 s_2 s_3$, $U$ is the potential energy of
the system, $p$ is the external pressure, $V_0$ the volume of the reference 
super-cell, and $N$ the total number of atoms in the system. Throughout all 
simulations described in this paper, the external pressure was set to zero.

We come now to the description of the potential. We have used the force field 
developed for the PE crystal by Sorensen, Liau, Kesner and Boyd \cite{sb}, with 
several slight modifications of the bonded interaction. This consists of 
diagonal terms corresponding to bond stretching, angle bending, and torsions, 
as well as of off-diagonal bond-bond, bond-angle and various angle-angle terms. 
For convenience, we have changed the form of the expansion in bond angles, 
replacing the expressions $(\theta - \theta_0)$ in all terms by 
$(\cos\theta - \cos\theta_0)/(-\sin\theta_0)$.
The original form of the vicinal bend-bend interaction, having two different 
force constants, $k_T$ and $k_G$ for torsional angles close to trans and gauche
minima, respectively, is useful for ground state studies, but not for a 
finite-temperature simulation, where the torsional angles may continuously 
fluctuate from one minimum to another. The form of this interaction has 
therefore been changed into the one used in Ref.\cite{kdg}, 
$ k \cos\varphi (\cos\theta_1 - \cos\theta_0) (\cos\theta_2 - \cos\theta_0)$, 
where $\theta_1,\theta_2$ are bond angles, $\varphi$ is a torsional angle, and 
$k$ is a new force constant. The value of the latter constant was taken
to be $-k_T/\sin^2\theta_0$ for C-C-C-C torsions and $2 k_G/\sin^2\theta_0$
for C-C-C-H torsions, in order to reproduce the curvature of the potential
in the vicinity of the ground state equilibrium value of each torsional angle in
the PE crystal. For H-C-C-H torsions, which are in the ground state in both 
trans and gauche minima, we took for $k$ an approximate value of
$k = \left(\sqrt{k_G k_T/\cos {{\pi}\over{3}} \cos {\pi}} \, \right)/
\sin^2\theta_0$, which 
guarantees that both original force constants $k_T$ and $k_G$ are approximated 
in the vicinity of the respective minimum with the same error of about 8 \%.
At this point we comment on the torsional potential used. As the force field
\cite{sb} has been originally designed for ground-state studies, the explicit
torsional potential for all torsions contains only 
the term $\cos 3\varphi$, which yields zero energy difference between the trans 
and gauche minima in C-C-C-C torsions. The actual difference thus comes
just from the non-bonded interaction superposed over 1 -- 4 atoms, and is, 
according to Ref. \cite{boydpr}, about 340 K, which is a somewhat higher value
than the generally accepted one of about 250 K. However, the use of periodic 
boundary conditions
anyway inhibits the creation of conformational defects strongly, and therefore 
this difference is not likely to play an important role, at least in the
temperature range studied. 

Concerning the nonbonded interaction, we used a spherical cutoff of 6 $\AA$ for 
all pair interactions, which corresponded to interaction of a given atom with 
about 110 neighbors. The list of neighbors has been determined at the 
beginning of the simulation with respect to the reference structure, and kept 
fixed throughout the evolution of the system (topological interaction). 
The use of the topological interaction would clearly preclude the longitudinal
diffusion of the chains by creating an artificial energy barrier, which might 
be unrealistic for a study of short alkanes. However, since our main interest 
lies in the limit of very long chains, this approximation, common in the study
of crystals, is acceptable here, and brings an advantage of considerably 
speeding up the execution of the program. 
The reference structure was obtained by placing the ideal crystal structure, 
described at the beginning of this section, inside the reference box, taking 
for the setting angle $|\psi|$ and the internal chain parameters the values of 
$|\psi| = 43.0 ^{\circ}, r_{CC} = 1.536 \AA, r_{CH} = 1.09 \AA, \theta_{HCH} = 
107.4 ^{\circ}$, respectively. These values turned out to be to a good 
approximation close to their 
true average values throughout the whole temperature range of the simulation, 
thus proving the consistency. Because of the relatively low cutoff used, we 
have added the long-range corrections to the non-bonded energy
and diagonal components of the pressure tensor. These corrections have been
calculated for the reference structure in the static lattice 
approximation, and tabulated for a suitable mesh of scaling factors 
$s_1,s_2,s_3$. During the simulation, the values of the corrections 
corresponding to the instantaneous values of the scaling factors were
calculated from the table by means of three-dimensional linear interpolation.
We note here that our way of treating the non-bonded interaction is different
from that used in \cite{sb}, where the interaction of a given atom with two
neighboring shells of chains has been taken into account and no long-range 
corrections have been applied. 

The MC sampling algorithm for PE crystal was described in considerable detail 
in our previous papers \cite{rm,rm1}, and here we present it just briefly.
In addition to the volume moves, we used local moves to move the atoms and 
global moves to move the chains. In the local moves, the atoms of the crystal 
lattice were visited in sequential order, and different maximum displacements 
have been used for the atoms of carbon and hydrogen, reflecting the fact that 
a carbon atom has four covalent bonds while a hydrogen atom has just one. The 
typical value of the acceptance ratio for the local moves was kept close to 0.3,
corresponding at temperature $T = 100$ K to isotropic maximum displacements of 
0.03 $\AA$ and 0.06 $\AA$ for carbons and hydrogens, respectively. In the global
moves, displacements of the center of mass of a whole chain along all three
axes accompanied by rotation of the chain along a line parallel to the 
crystallographic $c$-direction
and passing through the center of mass of the chain were attempted. Choosing 
the fraction of global moves to be 30 \% in this study, the global and local 
moves were alternated at random, in order to satisfy the detailed balance 
condition. Once it was decided to perform a global move, all the chains of the 
super-cell were attempted to move in sequential order. For C$_{12}$ chains at 
$T = 100$ K, the maximum (anisotropic) displacements of the chains 
were $\Delta x_{max} = \Delta y_{max} = 0.11 \AA, \Delta z_{max} = 0.05 \AA$, 
the maximum rotation angle being $\Delta \psi_{max} = 11^{\circ}$. This 
choice of the parameters resulted in an acceptance ratio of about 0.18 for the 
global moves. We note here that the energy change associated with a rigid 
displacement or a rotation of a whole chain scales linearly with the length of 
the chain, and therefore for longer chains we reduced appropriately the 
parameters of the global moves in order to preserve the same acceptance ratio.
No attempt to optimize the maximal displacements or fraction of different kinds
of moves has been done in this study. 
Concerning the volume moves, we attempted a change of all three scaling factors
$s_1,s_2,s_3$ after each sweep over the lattice (MCS) performing local or global
moves. For the super-cell $2 \times 3 \times 6$ unit cells at $T = 100$ K, the 
maximum changes of the scaling factors used were 
$\delta s_1 = \delta s_2 = 0.0045, \delta s_3 = 0.0009$, where the different 
values reflected the anisotropy of the diagonal elastic constants 
$c_{11},c_{22},c_{33}$. This choice resulted in an acceptance ratio of 0.21 for 
the volume moves. 

We have simulated several different super-cell sizes. The smallest one consisted
of $2 \times 3 \times 6$ unit cells, contained 12 C$_{12}$ chains with a total
of 432 atoms, and was used for simulation of the system at seven different 
temperatures ranging from 10 to 300 K. To study chain-length-dependent 
finite-size effects, at 
$T = 100$ K and $T = 300$ K the simulation was performed also on super-cells 
consisting of 12 C$_{24}$ and C$_{48}$ chains, containing respectively 
$2 \times 3 \times 12$ and $2 \times 3 \times 24$ unit cells, while at 
$T = 200$ K only the latter system size was studied in addition to the smallest
one. At $T = 300$ K, a super-cell size $4 \times 6 \times 12$ unit cells, twice
as large as the smallest one in each spatial direction, was also used, to study
volume-related finite-size effects. Finally, at all four highest temperatures, 
$T = 300,350,400,450$ K, a super-cell with the longest, C$_{96}$ chains was 
used, consisting of $2 \times 3 \times 48$ unit cells and containing 3456 atoms.

As the initial configuration for the lowest temperature simulation for a given 
super-cell size we used the corresponding reference structure.
In course of the simulation, we made use of the final configuration of the run 
at a lower temperature when possible, and always equilibrated the system for at
least $2 \times 10^4$ MCS before averaging. Our statistics is based on the run 
length of $1.4 \times 10^6$ MCS per data point for the smallest, 432 atom 
system, 
and a run length decreasing linearly with the number of atoms for the larger 
systems.

We have calculated the specific heat at constant pressure $c_p$ from the 
enthalpy fluctuations. The pressure tensor was calculated by means of a standard
virial expression. In order to check the consistency of the simulation 
algorithm,
we also evaluated the kinetic energy from the corresponding virial expression.
During the averaging, the accumulators for the total energy, lattice parameters 
$a,b,c$ and setting angle $\psi$ were updated after each MCS while those for 
virial and other
quantities were updated only every 10 MCS. Histograms have been accumulated for
the structural quantities, like the bond lengths and angles, torsional angles 
and the setting angle of the chains as well as the displacement of the center 
of mass of the chains along the $z$-axis. The whole run was always subdivided
into four batches and the batch subaverages were used to estimate the 
approximate error bars of the total averages. 

Concerning the elastic constants $c_{11}, c_{22}, c_{33}, c_{12}, c_{13}, 
c_{23}$ (in the Voigt's notation), we have independently determined them in two
different ways. Apart from the standard Parrinello-Rahman fluctuation formula 
\cite{pr}
\bqq
c_{ik} = {{k_B T}\over{\langle V \rangle}} \langle e_i e_k \rangle^{-1}   \; ,
i,k = 1,2,3 \; ,
\label{prff}
\eqq
we applied also the new fluctuation formula proposed in Ref.~\cite{gzs},
in its approximate version suitable for small strain fluctuations 
\bqq
c_{ik} = -\sum_n \langle p_i e_n \rangle \langle e_n e_k \rangle^{-1}   \; ,
\label{newff}
\eqq
where $p_i$ and $e_i$ are the diagonal components of the pressure tensor and
strain tensor, respectively.

\section{Results and discussion}

In this section we describe and discuss the results of the simulation. We start
with a comment on the stability of the crystal structure. The initial structure 
with the characteristic "herringbone"
arrangement of the chains was found to be stable at all temperatures and all 
system sizes except for the smallest system with C$_{12}$ chains, where an 
occasional rotation of a whole chain was observed at $T = 300 $ K. In the 
largest system with C$_{96}$ chains, no change of structure was observed up to 
$T = 450 $ K. Although the latter temperature is larger than the experimentally 
known melting temperature of PE crystal (414 K), the use of periodic 
boundary conditions inhibits the melting and allows the simulation of a 
superheated crystalline phase. On the other hand, our arrangement with 
constrained angles of the super-cell is compatible both with the orthorhombic 
and with the hexagonal phase, and would not prevent the system from entering 
the latter, which would occur when the ratio of the lattice parameters 
${{a}\over{b}}$ reaches the value $\sqrt{3}$. Our observation thus agrees with 
the experimentally known
stability of the orthorhombic structure up to the melting point. In order to
appreciate the amount of disorder present in the system at $T = 400$ K, we
show in Fig.1 (a) a projection of the atoms on the $xy$ plane for a typical 
configuration. The "herringbone" arrangement of the chains is still clearly
visible, in spite of a well pronounced disorder. In Fig.1 (b), a projection of
the same configuration on the $xz$ plane is shown. 

In Figs.2,3,4 we show the temperature dependence of the lattice parameters
$a,b,c$. Extrapolating these curves down to $T=0$, we find the ground state 
values of $a = 7.06 \AA,
b = 4.89 \AA, c = 2.530 \AA$. We note here that these values do not quite 
agree with those reported in Ref.\cite{sb}, where the following values 
have been found $a = 7.05 \AA, b = 4.94 \AA, c = 2.544 \AA$. We attribute these
discrepancies to the different way of treating the non-bonded interaction 
as well as to our slight modifications of the bonded interaction.

	Before discussing the thermal expansion of the lattice parameters,
we comment on the finite-size effects observed. A particularly pronounced one is
observed on the lattice parameter $b$, where the values for the smallest system
with C$_{12}$ chains are already at $T = 100$ K slightly larger than those for 
both systems with longer chains. The effect becomes stronger with increasing 
temperature. At $T = 300$ K, where we have data for four different chain 
lengths, the value of $b$ is clearly seen to increase with decreasing chain 
length, most dramatically for the system with C$_{12}$ chains. On the other 
hand, a considerably weaker finite-size effect is seen on lattice parameters
$a$ and $c$. While for the latter one it is perhaps not surprising, because of
the large stiffness of the system in the chain direction, the distinct behavior
of the $b$ parameter with respect to the $a$ parameter does not appear to be
so straightforward to interpret. We believe that its origin lies in the 
particular character of the chain packing, where the shortest hydrogen-hydrogen
contact distance is just that along the crystallographic $b$-direction 
($y$-axis) \cite{bookwund}.
This results in a stronger coupling of the lateral strain $\epsilon_2$ 
along the $b$-direction to the longitudinal displacements of whole chains. 
Since the 
fluctuations of these displacements are larger for systems with short chains, 
a particular finite-size effect arises. 

Concerning the thermal expansion itself, it is convenient to discuss separately
the case of the lateral lattice parameters $a,b$, and that of the axial one $c$.
We start with the lateral ones, and show in Fig.5 also the temperature 
dependence
of the aspect ratio ${{a}\over{b}}$. Two regimes can be clearly distinguished 
here. For temperatures up to about 250 K, both lattice parameters expand in 
a roughly
linear way with increasing temperature, and the ratio ${{a}\over{b}}$ raises
only slightly from its ground state value of 1.44 (which differs substantially 
from the value of $\sqrt{3} = 1.73$, corresponding to hexagonal structure). 
It is characteristic for this regime that the thermal expansion arising due to 
lattice anharmonicities can be described within a {\em phonon picture} using 
a quasi-harmonic or self-consistent quasi-harmonic approximation 
\cite{rutledge,hagele,stobbe}. For higher temperatures, the picture changes. 
While the lattice parameter $a$ starts to increase faster, the expansion of the
parameter $b$ at the same time 
becomes more slow until it develops a maximum at $T = 350$ K, where it starts 
to decrease again. Such behavior of $b$ has been already observed in the work 
\cite{rk}. As a consequence, the aspect ratio ${{a}\over{b}}$ increases
strongly. This suggests that the driving force of the change of the lateral 
lattice
parameters in this regime is the approach of a phase transition to a hexagonal 
phase, in which each chain is surrounded by six chains at equal distance. We 
have actually continued our simulations to even higher temperatures, and from
a limited amount of simulation performed in that region we have found an 
indication that the hexagonal phase is indeed reached in the range of 
temperatures 500 -- 550 K (Fig.5). 
It is, however, clear that in order to obtain reliable results from the 
simulations in this high-temperature range, where the phase transition in a
real PE crystal involves
large amplitude displacements and rotations of the whole chains, as well as a
considerable population of conformational defects, it would be 
necessary to introduce several modifications into the simulation algorithm.
We shall come back to this point again in the final section. 

In Fig.6, the temperature dependence of the average setting angle 
$\langle |\psi| \rangle$ of the chains 
is shown. It also fits well within the two regimes scenario, being rather flat 
up to $T = 300$ K, and then starting to decrease. This decrease
could indicate an approach of the value of $|\psi| = 30^{\circ}$, 
compatible with the symmetry of the hexagonal phase. It is also interesting to
note the pronounced finite-size effect, similar to that observed in the case of
$b$. 

Before discussing the temperature dependence of the axial lattice parameter
$c$, it is convenient to plot also the thermal expansion coefficients, defined 
as $\alpha_i = {a_i}^{-1} da_i/dT, i = 1,2,3$, where $a_1,a_2,a_3$ are the 
lattice parameters $a,b,c$. These have been obtained by taking the finite 
differences of the 
lattice parameters and are shown in Figs.7,8,9 as a function of temperature.
As their behavior trivially follows from that of the lattice parameters, 
discussed for $i = 1,2$ above, we just note here that all three coefficients 
converge at low temperatures to nonzero finite values, as can be expected in 
the classical limit. 

Concerning the behavior of $c$ and $\alpha_3$, the characteristic feature is 
that $\alpha_3$ is negative in the whole temperature range and an order of 
magnitude smaller than $\alpha_1$. It is interesting here to compare our result
for $\alpha_3$ to that found in Ref. \cite{rutledge} within the quasi-harmonic 
approximation for a different force field \cite{kdg}. 
A distinct feature of the latter result is that 
the classical value of $\alpha_3$ is considerably smaller in magnitude than the 
quantum mechanical value (and also the experimental one), and approaches 
zero as $T \to 0$. Since in the classical limit there is no a priori
reason for such behavior, it has to be regarded as accidental. According to 
the argumentation in Ref. \cite{rutledge}, there are contributions of different
sign to $\alpha_3$ from different phonon modes, negative from lattice modes
(mainly backbone torsions), and positive from the harder ones. In the classical
limit, all the modes contribute at all temperatures and happen to 
just cancel each other as $T \to 0$. In order to estimate the amount of
contribution of the torsions in our case, we have made use of the work 
\cite{hagele}, where a formula is derived for the axial thermal contraction in 
a simple one-chain model with only torsional degrees of freedom. In the 
classical limit, the formula predicts $c - c_0 = - {{1}\over{4}} c_0 
\sin^2 {{\alpha}\over{2}} \langle \phi_{CCCC}^{2} \rangle$, where 
$\alpha = \pi - \theta_{CCC}$, and $\phi_{CCCC}$ is the fluctuation of the
torsional angle $\varphi_{CCCC}$ from the trans minimum. In Fig.10 we have 
plotted
$c$ against $\langle \phi_{CCCC}^{2} \rangle$, and found a very good linear 
dependence over
the whole temperature range up to 450 K, where the effective torsional constant,
defined as $C_{tors} = {{T}\over{\langle \phi_{CCCC}^2 \rangle}}$, undergoes 
a considerable softening with increasing temperature (Fig.11). The 
proportionality constant determined from our plot was, however, larger 
in magnitude by about 50 \% with respect to the above
value of ${{1}\over{4}} c_0 \sin^2 {{\alpha}\over{2}}$, valid for the simple
model. The linear dependence suggests that with the force field used \cite{sb},
the negative $\alpha_3$ originates almost entirely from the C-C-C-C torsions, 
which
points to a certain intrinsic difference in the anharmonic properties of the 
force fields \cite{sb} and \cite{kdg}. The pronounced increase of $\alpha_3$ 
for $T > 300$ K
is thus a consequence of a strong softening of the torsional potential due to 
the large amplitude of the fluctuations.

For comparison of these results to experimental ones, we have chosen two sets of
data. In the work \cite{davis}, lattice parameters $a,b,c$ have been measured 
in the temperature range 93 -- 333 K, and the data are smooth enough to allow
a direct extraction of thermal expansion coefficients by means of finite 
differences. The other chosen set of data \cite{sl} is to our knowledge 
the only one covering the range from helium temperatures up to $T = 350$ K, 
however, the scatter of the data is too large for a direct numerical 
differentiation. Therefore we decided to first fit them by a fourth-order 
polynomial and then evaluate the expansion coefficients analytically. For both 
sets, the experimental values of the lattice parameters are shown in Figs.2,3,4
and the corresponding thermal expansion coefficients in Figs.7,8,9 . 

Concerning the absolute value of the lattice parameters, in case of $a$ our 
results agree well with the data \cite{davis}, while falling slightly below the
data \cite{sl}, in particular at the lowest temperatures. In case of $b$ our 
results fall slightly above and in that of $c$ slightly below both data sets in 
the whole temperature range. As far as the temperature dependence itself is
concerned, this is most conveniently discussed in terms
of the thermal expansion coefficients $\alpha_i$. We note first that for 
temperatures lower
than about 150 K, quantum effects become crucial and cannot be neglected, as 
they are responsible for the vanishing of all expansion coefficients in the 
limit 
$T \to 0$. A meaningful comparison of our classical results to experimental data
is thus possible only for larger temperatures. For $T > 150$ K, $\alpha_1$ 
agrees qualitatively well with the data \cite{davis}, although the experimental
ones appear to be slightly smaller. In particular, the pronounced increase of 
$\alpha_1$ for $T > 250$ K appears to be well reproduced, in contrast to the 
result obtained in Ref.\cite{rutledge} within the classical quasi-harmonic 
approximation for the force field \cite{kdg}. The agreement is less good for 
the data \cite{sl}, which are distinctly smaller and start to increase strongly
at somewhat smaller temperatures, for $T > 200$ K. In case of $\alpha_2$, our
data agree for $T > 150$ K qualitatively well with the set \cite{sl}, correctly
reproducing the gradual decrease with temperature, although our results are 
somewhat smaller. The data \cite{davis} exhibit here a different behavior,
being of the same magnitude as the ones in Ref. \cite{sl}, but markedly flat in
the whole range of temperatures. We note here that in the work \cite{swan}, 
$\alpha_2 < 0$ has been experimentally observed just below the melting point.
On the other hand, the classical result for $\alpha_2$ calculated in 
Ref.\cite{rutledge} exhibits instead an upward curvature. Concerning 
$\alpha_3$, our results agree quantitatively well with the data \cite{davis}
in the whole range of temperatures, although some scatter of the data precludes
here a more detailed comparison. In the set \cite{sl}, $\alpha_3$ behaves for 
$T > 100$ K qualitatively similar to our results, however, the plateau value is
somewhat smaller and the strong increase in magnitude appears at a lower 
temperature, already between 200 - 250 K. The origin of some of the observed 
discrepancies may be either in the force field itself, or in the quantum 
effects, which may be relevant even at temperatures as high as 300 K
\cite{rutledge}. Also our fit of the data \cite{sl} is not unique, and may
itself be a source of additional errors in the coefficients $\alpha_i$. 
Last, but not least, it appears that there exists also a considerable scatter 
between the experimental data from different sources. We therefore believe that
it would be interesting to 
re-examine the temperature dependence of the structural parameters (including 
possibly the setting angle of the chains) of well-crystalline samples of PE, 
in the whole range from very low temperatures up to the melting point,
using up-to-date X-ray or neutron diffraction techniques. 

In Fig.12, the average fluctuations $\sqrt{\langle (\delta |\psi|)^2 \rangle}$ 
of the setting angle of the chains are shown as a function of temperature. Apart
from a trivial finite-size effect of average fluctuation increasing with 
decreasing chain length, we note that for the system with C$_{12}$ chains
the curve exhibits a marked enhancement of the fluctuations at $T = 300$ K. 
Such enhancement suggests that the system is approaching a transition into a 
phase similar to the "rotator" phases of $n$-paraffins, where the setting angle
of the chains jumps among several minima. For the system with C$_{96}$ chains, 
the same phenomenon occurs only for $T > 400$ K, which in a real PE crystal 
coincides with the melting point. In Fig.13, we show a histogram of the setting
angle $\psi$ for the system with C$_{96}$ chains at the highest temperature 
$T = 450$ K. The distribution is still bimodal, with two peaks centered at 
$\pm \langle |\psi| \rangle$, as it is in the ground state,
corresponding to the "herringbone" arrangement of the chains. This proves that 
in our super-cell with periodic boundary conditions in all directions, the 
superheated orthorhombic structure is stable even at such high temperature, at 
least on the MC time scale of our run.

We comment now briefly on some other internal structural parameters. In Figs.14
and 15, we show the average angles $\langle \theta_{CCC} \rangle$ and 
$\langle \theta_{HCH} \rangle$ as a function of temperature. While the angle 
$\langle \theta_{CCC} \rangle$ develops between 250 and 300 K a very shallow 
minimum, 
the angle $\langle \theta_{HCH} \rangle$ decreases roughly linearly throughout 
the whole range of temperatures, its overall variation being about a factor of 3
larger than that of the former. Both average bond lengths 
$\langle r_{CC} \rangle$ and $\langle r_{CH} \rangle$ increase very slightly 
with temperature, the dependence being very close to linear, 
$\langle r_{CC} \rangle$  varying from 1.5357 $\AA$ at 10 K to 1.5398 $\AA$ at 
450 K, and $\langle r_{CH} \rangle$  varying from 1.0898 $\AA$ at 10 K to 
1.0924 $\AA$ at 450 K. It is interesting to show the average torsional 
fluctuations $\sqrt{\langle \phi_{CCCC}^2 \rangle}$
as a function of temperature, Fig.16, where we see an enhancement
for $T > 400$ K, corresponding to already discussed softening of the effective
torsional potential (Fig.11). The typical fluctuation in this region is about 
$13^{\circ}$. From the histograms of the torsional angles we found that for
temperatures up to 400 K, no gauche defects are created in the chains, while at
450 K only an extremely low population starts to arise. This is clearly a 
consequence of the periodic boundary conditions used in the chain direction
together with the still relatively short length of the chains used.

We come now to some thermodynamic parameters, and start with the specific heat
per unit cell. This is shown in Fig.17 as a function of temperature. At low 
temperatures, where the classical crystal is always harmonic, 
${{c_p}\over{k_B}}$ reaches the classical equipartition value of 18, 
corresponding to
$3 \times 12 = 36$ degrees of freedom per unit cell. With increasing 
temperature, as the anharmonicities become important, its value gradually
increases, until a finite size effect starts to appear for $T > 200$ K. While 
for the system with C$_{12}$ chains $c_p$ starts to grow faster for 
$T > 200$ K, for the system with C$_{96}$ chains this occurs only at about
300 K. This behavior is likely to be related to the already discussed 
chain-length-dependent onset of increased fluctuations of the setting angle, 
or rotations of the chains around their axes. 

In Figs.18 -- 22, we show the elastic constants $c_{11},c_{22},c_{12},c_{13},
c_{23}$, determined by means of the Parrinello-Rahman fluctuation formula 
(\ref{prff}), as a function of temperature. For the case of $c_{33}$, we show 
in Fig.23 both the values obtained according 
to the Parrinello-Rahman fluctuation formula (\ref{prff}) and those found from 
the new formula (\ref{newff}). We see that the values obtained with the latter 
one have considerably smaller scatter. Actually, we have tried to use the 
formula (\ref{newff}) also for other elastic constants, however, only in the
case of $c_{33}$ a definite improvement of the convergence was observed. 
It would certainly be desirable to be able to improve on the accuracy of the 
elastic constants, in particular in case of $c_{13},c_{23}$; however, at present
this seems to require a prohibitively large CPU time unless a considerably more
efficient algorithm is available. At this point we note that some of the error 
bars shown in the figures for the elastic constants are probably an
underestimate of the true ones, as they have been obtained from the variance of
the results from just four batches.

To discuss now
the temperature dependence of the elastic constants, it's again convenient to
do so separately for the lateral ones $c_{11},c_{22},c_{12}$, and for the ones
related to the axial strain $\epsilon_{3}$, namely $c_{13},c_{23}, c_{33}$. All
three constants $c_{11},c_{22},c_{12}$ are seen to monotonously decrease with 
temperature, reaching at 400 K, just below the melting point, about 60 \%
or even less of their ground state values. This behavior originates mainly 
from the thermal expansion of the crystal in
the lateral directions and is typical for van der Waals systems, like, e.g.,
solid argon \cite{loeding}. An interesting finite-size effect is seen on
the diagonal elastic constant $c_{22}$, which is at $T = 300$ K distinctly 
smaller for the system with C$_{12}$ chains than for other systems. This 
observation correlates with the effect seen on lattice parameter $b$, which was
in turn found to be larger for the system with the shortest chains. 

Concerning the other group of elastic constants, we first note that the diagonal
one $c_{33}$ is two orders of magnitude larger than all the other elastic 
constants. At low temperatures, $c_{33}$ reaches a limiting value as large as 
340 GPa and decreases monotonously with increasing temperature, dropping at 
400 K to about 80 \% of its ground state value. On the other hand, the two 
off-diagonal elastic constants, $c_{13}$ and $c_{23}$, are, interestingly, 
found to increase with temperature, in agreement with the results in 
Ref.\cite{rutledge}. Taking into account the negative thermal expansion of the 
system in the axial direction, the behavior of this group of elastic constants
does not appear to be just a trivial consequence of the thermal expansion, as
it was in the case of $c_{11},c_{22},c_{12}$. In order to understand such 
behavior, it is interesting to plot also the elastic
constant $c_{33}$ vs. the mean-square fluctuation of the torsional angle
$\langle \phi_{CCCC}^{2} \rangle$, Fig.24. We find again a roughly linear 
dependence (compare Fig.10), which
suggests that the softening of $c_{33}$ is directly related to the activation of
the torsional degrees of freedom. The mechanism underlying the behavior of 
these elastic constants then might be the following. At $T = 0$, 
where the chain backbones are perfectly flat in their all-trans states, an axial
deformation can be accommodated only by bending the bond angles $\theta_{CCC}$. 
Angle bending is, however, after bond stretching, the second most stiff degree 
of freedom in the 
system and determines the high ground state value of $c_{33}$. As the torsional 
fluctuations become activated with increasing temperature, the chains start to 
"wiggle" and develop 
transverse fluctuations (see Fig.1 (b)). In such configurations, it becomes
possible to accommodate a part of the axial deformation in the torsional modes, 
which represent the most soft ones in the bonded interaction, and $c_{33}$ is
therefore renormalized to a smaller value. This must be, however, accompanied 
by an increased response of the transverse fluctuations of the chains, resulting
in lateral strain, because of the 
"wiggling", since the total length of the chains is very hard to change. 
An increased lateral strain response to an axial strain means, however, just
an increase of the value of the elastic constants $c_{13}$ and $c_{23}$. 
It would be, of course, desirable to have a quantitative theory supporting this
intuitive, but, as we believe, well plausible interpretation.

The reason why the constants $c_{13}$ and $c_{23}$ appear to have the largest
error bars among all elastic constants also seems to be connected with the fact
that these two ones express just the coupling between the lateral and axial 
strains. The computational efficiency of the algorithm in determination of 
these two quantities is crucially dependent on the exchange of energy
between the non-bonded and bonded interactions, which is probably still the most
difficult point even with the present algorithm, because of the large separation
of the relevant energy scales.

Before closing this section, we would like to make yet few more remarks on the 
finite-size effects. Comparing the results at $T = 300$ K for both systems 
consisting of C$_{24}$ chains, containing respectively $2 \times 3 \times 12$ 
and $4 \times 6 \times 12$ unit cells, we see that the values obtained with both
system sizes are for all quantities practically equal. This suggests that the 
finite-size effects related directly to the volume of the box are relatively 
small, at least for the two system sizes considered (which are, however, still 
not very large). On the other hand, a definite chain-length-dependent 
finite-size effect has been found in case of several quantities for the chain 
lengths 
considered, its magnitude being also distinctly temperature dependent. For 
practical purposes, it results that the smallest system size used with C$_{12}$
chains can be representative of a classical PE crystal only at rather low 
temperatures, perhaps below 100 K, since at and above this temperature it 
exhibits pronounced finite-size effects.
On the other hand, the systems with C$_{24}$ and C$_{48}$ chains appear to 
represent the classical PE crystal reasonably well for temperatures lower than 
300 K, while at this and higher temperatures the use of a system with C$_{96}$ 
chains or even longer would be strongly recommended.

\section{Conclusions}

In this paper, we have demonstrated three main points. First, the MC algorithm 
using global moves on the chains in addition to the local moves on the atoms is
a well applicable method for a classical simulation of crystalline PE. It allows
an accurate determination of 
the structural properties and yields also fairly accurate results for the 
elastic constants. Second, the force field we have used \cite{sb} is well
able to reproduce the experimentally known structure in the whole range
of temperatures, and where the classical description is appropriate, the results
obtained agree well with the available experimental data. Third, we have studied
the finite-size effects, mainly due to chain length, and determined for 
different temperatures a minimal chain length necessary for the system to be in
the limit of long chains, and thus representative of PE. All these findings look
promising for further studies, and here we suggest some possible directions. 

Basically, there are two routes to extend the present study, concentrating on 
the low-temperature and high-temperature region, respectively. The first one
would aim on taking into account the quantum effects, e.g. by means of a Path 
Integral MC technique. Apart from improving the agreement with experiment, 
mainly
at low temperatures, this technique is able to treat the quantum effects at a
finite temperature in an essentially exact way and therefore should also allow 
to check the range of validity of various approximate treatments, like the 
quasi-harmonic or self-consistent quasi-harmonic approximations 
\cite{rutledge,hagele}. This would help to understand better the true importance
of quantum effects at different temperatures in this paradigmatic crystalline 
polymer system.

The second route would aim on the high-temperature region of the phase diagram,
where the orthorhombic crystal melts under normal pressure, but is known to 
undergo a phase transition into a hexagonal "condis" phase at elevated pressure 
\cite{condis}. In the present simulation arrangement, melting is prevented by
periodic boundary conditions in both lateral directions, but the same boundary 
conditions being applied in the chain direction inhibit also the creation of 
conformational defects. Nevertheless, there are indications arising from the 
present simulation that a similar transition could indeed occur in the 
temperature 
region 500 -- 550 K, the main ones being the approach of the aspect ratio 
${{a}\over{b}}$ towards the hexagonal value of $\sqrt{3}$, and enhancement of 
the torsional angle and setting angle fluctuations for temperatures over 400 K.
In order to study this high-temperature regime properly, several modifications
of the present algorithm would be necessary. The main one would be a lifting of
the periodic boundary conditions in the chain direction, thus introducing free 
chain ends. A use of a larger cutoff for the non-bonded interactions would be 
necessary in order to treat correctly the large amplitude displacements and 
rotations of the chains involved in the transition, and perhaps a non-spherical
cutoff including certain number of atoms from one or two neighboring shells of
chains around a given atom might be a preferred solution.
It would also be necessary to modify the torsional potential in order to yield
a correct value for the energy difference between the trans and gauche states.
In order to be able to sample configurations with a considerable population of 
conformational defects, it might also be useful to introduce some other kind of
MC moves acting on the torsional degrees of freedom. Finally, a full MC version
of the Parrinello-Rahman variable-cell-shape technique should be introduced in 
order
to allow also for shear fluctuations. These might in principle be substantially
involved in the transition itself, although the hexagonal phase can be reached 
from the orthorhombic one without creating a static shear strain.

Before closing, we also emphasize that techniques similar to those applied
here should be useful for a wide variety of other macromolecular crystals. 

\acknowledgements

We would like to acknowledge stimulating discussions with 
A. A. Gusev, P. C. H\"{a}gele, K. Kremer, R. J. Meier, A. Milchev, 
F. M\"{u}ller-Plathe, M. M\"{u}ser,
P. Nielaba, G. C. Rutledge, G. Smith, U. W. Suter, E. Tosatti, M. M. Zehnder, 
as well as correspondence with R. H. Boyd and R. A. Stobbe.

\begin{figure}
Fig.1. A snapshot of a projection of a typical configuration of the system
at $T = 400$ K on the $xy$ plane (a), and on the $xz$ plane (b). The size of 
the reference super-cell is $L_x = 14.5 \AA, L_y = 15.0 \AA, L_z = 121.44 \AA$.
Filled points represent the carbon atoms, empty ones the hydrogen atoms.
Note that the "herringbone" arrangement of the chains is still clearly visible
in (a).
\end{figure}

\begin{figure}
Fig.2. Temperature dependence of the lattice parameter $a$, shown for different
system sizes, together with the experimental data \cite{davis,sl}. In this and 
most of the following figures, statistical error bars are shown, lines are for 
visual help only.
\end{figure}

\begin{figure}
Fig.3. Temperature dependence of the lattice parameter $b$, shown for different
system sizes, together with the experimental data \cite{davis,sl}. Note the 
finite-size effect.
\end{figure}

\begin{figure}
Fig.4. Temperature dependence of the lattice parameter $c$, shown for different
system sizes, together with the experimental data \cite{davis,sl}.
\end{figure}

\begin{figure}
Fig.5. Temperature dependence of the aspect ratio ${{a}\over{b}}$. Note the 
approach towards the hexagonal value of $\sqrt{3} = 1.732$ with increasing 
temperature. 
\end{figure}

\begin{figure}
Fig.6. Temperature dependence of the average setting angle 
$\langle |\psi| \rangle$ of the chains, shown for different system sizes.
Note the finite-size effect.
\end{figure}

\begin{figure}
Fig.7. Thermal expansion coefficient $\alpha_1$ as a function of temperature,
shown for different system sizes, together with the experimental data 
\cite{davis,sl}.
\end{figure}

\begin{figure}
Fig.8. Thermal expansion coefficient $\alpha_2$ as a function of temperature,
shown for different system sizes, together with the experimental data 
\cite{davis,sl}. Note the finite-size effect.
\end{figure}

\begin{figure}
Fig.9. Thermal expansion coefficient $\alpha_3$ as a function of temperature,
shown for different system sizes, together with the experimental data 
\cite{davis,sl}.
\end{figure}

\begin{figure}
Fig.10. Lattice parameter $c$ vs. $\langle \phi_{CCCC}^2 \rangle$, shown for 
different system sizes. Note
the linear dependence over the whole temperature range, which can be fitted
by a line $c = 2.530 - 9.50 \times 10^{-5} \langle \phi_{CCCC}^2 \rangle$. 
\end{figure}

\begin{figure}
Fig.11. Temperature dependence of the effective torsional force constant 
$C_{tors} = {{T}\over{\langle \phi_{CCCC}^2 \rangle}}$, shown for different 
system sizes. Note the considerable softening with increasing temperature.
\end{figure}

\begin{figure}
Fig.12. Temperature dependence of the average fluctuation 
$\sqrt{\langle (\delta |\psi|)^2 \rangle}$ of the setting angle of the chains, 
shown for different system sizes.
\end{figure}

\begin{figure}
Fig.13. Histogram of the setting angle $\psi$ of the chains for the system with
C$_{96}$ chains at the highest temperature $T = 450$ K. Note that the 
distribution is still bimodal, corresponding to the "herringbone" arrangement
of the chains. 
\end{figure}

\begin{figure}
Fig.14. Temperature dependence of the average bond angle 
$\langle \theta_{CCC} \rangle$, shown for different system sizes. Note the 
minimum at 300 K.
\end{figure}

\begin{figure}
Fig.15. Temperature dependence of the average bond angle 
$\langle \theta_{HCH} \rangle$, shown for different system sizes. 
\end{figure}

\begin{figure}
Fig.16. Temperature dependence of the average fluctuation
$\sqrt{\langle \phi_{CCCC}^2 \rangle}$ of the torsional angle $\varphi_{CCCC}$ 
from the trans minimum, shown for different system sizes. 
\end{figure}

\begin{figure}
Fig.17. Temperature dependence of the specific heat per unit cell at constant 
pressure $c_p$, shown for different system sizes.
\end{figure}

\begin{figure}
Fig.18. Elastic constant $c_{11}$ as a function of temperature, shown for 
different system sizes.
\end{figure}

\begin{figure}
Fig.19. Elastic constant $c_{22}$ as a function of temperature, shown for 
different system sizes. Note the finite-size effect.
\end{figure}

\begin{figure}
Fig.20. Elastic constant $c_{12}$ as a function of temperature, shown for 
different system sizes.
\end{figure}

\begin{figure}
Fig.21. Elastic constant $c_{13}$ as a function of temperature, shown for 
different system sizes. Note the increase with temperature.
\end{figure}

\begin{figure}
Fig.22. Elastic constant $c_{23}$ as a function of temperature, shown for 
different system sizes. Note the increase with temperature.
\end{figure}

\begin{figure}
Fig.23. Elastic constant $c_{33}$ as a function of temperature, shown for
different system sizes. Empty and filled symbols represent data determined from 
the formulas (\ref{prff}) and (\ref{newff}), respectively. Note the smaller 
scatter of the latter results.
\end{figure}

\begin{figure}
Fig.24. Elastic constant $c_{33}$ vs. $\langle \phi_{CCCC}^2 \rangle$, shown for
different system sizes. Note the roughly linear dependence.
\end{figure}

\end{document}